\documentclass[aps,pre,showpacs,amsmath,amssymb]{revtex4}
\usepackage{graphicx}

\newcommand{\bs}[1]{{\boldsymbol #1 }}
\newcommand{\ave}[1]{\langle #1 \rangle}

\usepackage{graphicx}
\usepackage{bm}

\begin{document}

\preprint{}

\title{Nonequilibrium mesoscopic Fermi reservoirs distributions and particle current through a coherent quantum system}
\author{Shigeru Ajisaka}
\affiliation{Departamento de F\'isica, Facultad de Ciencias F\'isicas y Matem\'aticas, Universidad de Chile, Casilla 487-3, Santiago Chile}
\author{Felipe Barra}
\affiliation{Departamento de F\'isica, Facultad de Ciencias F\'isicas y Matem\'aticas, Universidad de Chile, Casilla 487-3, Santiago Chile}

\date{\today}% It is always \today, today,
\begin{abstract}
  We study particle current and occupation distribution in a recently proposed model for coherent quantum transport.
In this model a system connected to mesoscopic Fermi reservoirs (mesoreservoir) is driven out of
equilibrium by the action of superreservoirs with prescribed temperatures and
 chemical potentials described by a simple dissipative mechanism with the
Lindblad equation.  We compare exact (numerical) results for the non-equilibrium steady state particle current with theoretical expectations based on the Landauer formula
and show that the model reproduce the behavior of coherent quantum systems  in the expected parameter region. We also obtain the occupation distribution on the mesoreservoir in the non-equilibrium steady state and 
compare them
with the occupation distribution on the leads in usual description of coherent quantum transport. 
\end{abstract}

%03.65.Fd Algebraic methods in quantum mechanics
%05.30.Fk Fermions systems (quantum statistical mechanics)
%05.60.Gg Quantum transport
%05.70.Ln Nonequilibrium and irreversible thermodynamics

\maketitle

\section{introduction}
The current of non-interacting particles flowing from a left to a right hand side reservoir trough a coherent mesoscopic conductor is usually described by the ``Landauer picture" :
Electrons in the left (right) reservoir that are Fermi distributed with chemical potential $\mu_{L}$ ($\mu_{R}$) and inverse temperature $\beta_{L}$ ($\beta_{R}$) can come close to the conductor and feed a scattering state that can transmit it to the right (left) reservoir. All possible dissipative processes such as thermalization occur in the reservoirs while the system formed by the conductor and the leads is assumed to be coherent. 
The probability of being transmitted is  a property of the conductor connected to the leads, which is treated as a scattering system. 
In this picture, the probability that an outgoing electron comes back to the conductor before being thermalized is neglected, 
the contact is said to be reflectionless~\cite{DattaBook, imry}. 

In the linear regime (small chemical potential difference) this scenario predict that $G=\frac{e^2}{2\pi\hbar}T(\epsilon_F)$ is the conductance, where $e$ is the electron charge, $2\pi\hbar$ is Plank«s constant and $T(\epsilon_F)$ the total transmission probability evaluated at the Fermi level~\cite{imry}. For simplicity we will consider spineless fermions but the generalization to the spinfull case is straightforward, in that case a factor two should be added to the conductance. For a perfect one-dimensional conductor $T(\epsilon_F)=1$ this formula predict a finite resistance. Instead Landauer~\cite{Landauer1957} suggested that the conductance should be $G_0=\frac{e^2}{2\pi\hbar}T(\epsilon_F)/R(\epsilon_F)$ with $R(\epsilon_F)$ the reflection coefficient that vanish for a perfect conductor reestablishing the intuitive idea that an ideal conductor should have zero resistance. There was a long controversy regarding the validity of each formula that was finally settled with the realization of unavoidable contact resistance $1/G_c$, which is such that $1/G_c +1/G_0+1/G_c=1/G$. Associated to this contact resistance there is a potential drop~\cite{Pinhas1985,DattaBook} taking place between the reservoir and the lead. Note that the resistance are added in series , therefore it is assumed that the contacts are not part of the coherent system. 
 
Recently several works have shown that the Landauer formula for the conductance or more generally the averaged particle current follows from 
a first principle description of the non-equilibrium steady state (NESS) of the non-interacting many-particle conductor-leads plus reservoirs system.    
In some particular limiting situations such as infinite reservoirs, this has been done rigorously~\cite{needref} or using powerful tools such as non-equilibrium green function formalism that allows to treat the case where particles interact inside the system~\cite{meir}. Most of these works focus their attention 
in the conductor and current but no attention is paid to the NESS properties of the leads or reservoirs, which are usually assumed to be in equilibrium.
However, near the conductor deviations from equilibrium appear in reality as was 
recently observed in a cold-atom analog of a mesoscopic conductor~\cite{science}.

In this paper we would like to discuss the relation between the current through a conductor and the non-equilibrium distributions on the leads in the framework of a recently proposed model~\cite{Shigeru} for a system plus mesoscopic reservoir that physically represent the leads that couple the conductor to the superreservoirs. Dissipation is included in our model and affects only the leads.  The model is obtained by tracing out the superreservoir degrees of freedom in the unitary evolution  equation for the density matrix of the full reservoir plus conductor system. At the price of several approximations such as the Born-Markov approximation (see, e.g., 
Ref. ~\cite{Wichterich07, kosov2, BreuerBook}) a master equation in the Lindblad form for the reduced density operator of the conductor plus leads is obtained. The Lindblad dissipator encodes the effect of the Markovian superreservoir on the leads. 

The structure of the paper is as follows: In section \ref{sec.2} we briefly present the model, in section \ref{sec.current} we analyze the current flow and its agreement with Landauer formula and the occupation distribution on the mesoreservoirs in comparison with
expectations based on the Landauer picture. Finally in section \ref{conclu} we present our conclusions. 

%%%%%%%%%%%%%%%%%%%%%%%%%%%%%%%%%%%%%%%%%%%%%%%%%%%%%%%%%%%%%%%%%%%%%%

\section{Description of the model} 
\label{sec.2}
As our ``conductor" we consider both, a quantum dot and a one-dimensional quantum chain coupled at its left and right  boundaries to finite reservoirs, which hereafter we call mesoreservoirs. The Hamiltonian of the total system can be written
as $H = H_S+H_L+H_R+V$, where
\begin{equation}
\label{hs}
H_S =  U c^\dag c
\end{equation}
is the Hamiltonian for the quantum dot or 
\begin{equation}
\label{hs}
H_S = -\sum_{j=1}^{N-1} \Big(t_j c_j^\dag c_{j+1} + ({\rm h.c.}) \Big)
+\sum_{j=1}^{N} U_j c_j^\dag c_j
\end{equation}
is the Hamiltonian of the chain 
with $t_j$ the nearest neighbor hopping, $U_j$ the
onsite potential and  $c_j,c_j^\dagger$ the  annihilation/creation operator for the spinless fermions on the site $j$ of the chain (for the dot we do not need the subscript). The chain (or dot) interacts  with the mesoreservoirs through the term
\begin{equation}
V =\sum_{k=1}^{K} \left( v^L_{k}a_{kL}^\dag c_1 + v^R_{k}a_{kR}^\dag c_n\right) + ({\rm h.c.}),
\label{V1}
\end{equation}
(in the dot case $c_1$ and $c_n$ should be replaced by $c$). The mesoreservoirs Hamiltonian are $H_{\alpha}= \sum_{k=1}^{K} 
\varepsilon_k a_{k\alpha}^\dag a_{k\alpha}$,  here $\alpha\!=\!\{L, R\}$ denotes the left and right mesoreservoir.
They corresponds to a finite number of spinless fermions with wave number $k\in\{1,\ldots,K\}$
sharing the same spectrum with a constant density of states $\theta_0$ in the band $[E_{\rm min},E_{\rm max}]$ described by $\varepsilon_k \equiv \theta_0 (k-k_0)$ and  
$a_{k,\alpha},a^\dagger_{k,\alpha}$  are annihilation/creation operator  of the  $\alpha$ mesoreservoir.
The system is coupled to the mesoreservoirs only at the extreme sites 
of the chain with coupling strength $v_k^\alpha$ that we choose $k$-independent~\footnote{In general we can include couplings to deeper sites of the chain and also $k$ dependent super-reservoir to mesoreservoir couplings $\gamma_k$, see~\cite{mariborg}.} $v_k^\alpha = v_\alpha$. 

We assume that  the density matrix of the conductor plus mesoreservoirs evolves according to the 
many-body Lindblad equation (hereafter we consider units where $e=1$ and $\hbar=1$)
\begin{eqnarray}
\frac{{\rm d}}{{\rm d}t}\rho &=&
-i[H,\rho] 
+\sum_{k,\alpha,m}
\left(
2L_{k, \alpha, m} \rho L_{k, \alpha, m}^\dag
-\{L_{k, \alpha, m}^\dag L_{k, \alpha, m},\rho\}
\right),
\label{eq:lindbladmodel}
\end{eqnarray}
where $m\in\{1,2\}$ and $L_{k,\alpha,1} = \sqrt{\gamma ( 1 - f_{\alpha}(\varepsilon_k) )} a_{k\alpha},\quad L_{k,\alpha,2}=\sqrt{\gamma f_{\alpha} (\varepsilon_k)} a_{k\alpha}^\dag$ are operators  representing the coupling of the mesoreservoirs to the
superreservoirs, $f_\alpha(\varepsilon)=(e^{\beta_\alpha  (\varepsilon-\mu_\alpha)}+1)^{-1}$   
are Fermi distributions, with inverse   temperatures $\beta_\alpha$ and chemical potentials
$\mu_\alpha$, and  $[\cdot , \cdot]$  and $\{ \cdot ,  \cdot\}$ denote
the commutator and anti-commutator, respectively. 
The parameter $\gamma$ determines the strength of the coupling to the
superreservoirs and to keep the model as simple as possible we take it
constant.  
The form of the Lindblad dissipators is such that in the absence of coupling to the chain (i.e. $v_\alpha=0$), when the mesoreservoir is only coupled to the superreservoir, the former is in an equilibrium state described by Fermi distribution~\cite{prosen08,Kosov11}.  Physically the mesoreservoir mimic the leads (with a finite number of modes) that connect the macroscopic (infinite) reservoir to the system. This model has been justified starting from a Hamiltonian description of the baths in the case of the quantum dot~\cite{kosov2} in the limit $\gamma\to 0$, $K\to \infty$ and we apply it also for short chains ($N\ll K$).
%%%%%%%%%%%%%%%%%%%%%%%%%%%%%%%%%%%%%%%%%%%%%%%%%%%%%%%%%%%%%%%%%%%%%%

\section{Analysis and properties of the model}
\label{sec.current}
To analyze our model we use the formalism developed in
\cite{Prosen10}. There it is shown that the spectrum of the evolution superoperator is given
in terms of the eigenvalues $s_j$ (so-called rapidities) of a matrix $X$ which in our case is given by
\begin{eqnarray*}
\bs{X} &=& -\frac{\rm i}{2}\bs{H}\otimes \sigma_y + 
	\frac{\gamma}{2}
	\begin{pmatrix}
	\bs{E}_{K} & \bs{0}_{K\times N} &\bs{0}_{K\times K}\\
	\bs{0}_{N\times K} & \bs{0}_{n\times N} &\bs{0}_{N\times K}\\
	\bs{0}_{K\times K} & \bs{0}_{K\times N} &\bs{E}_{K}
	\end{pmatrix}
	\otimes\bs{E}_2,
\end{eqnarray*}
where $\bs{0}_{i\times j}$ and $E_j$ denote $i\times j$ zero matrix
and $j\times j$ unit matrix, $\sigma_y$ is the Pauli matrix, and
$\bs{H}$ is a matrix which defines the quadratic form of the
Hamiltonian, as $H= \bs{d}^\dagger \bs{H} \bs{d}$ in terms of fermionic
operators $\bs{d}^T \equiv \{a_{1L},\cdots
a_{KL},c_1,\cdots,c_N,a_{1R},\cdots a_{KR} \}$.

The NESS average of a quadratic observable like $d^\dagger_jd_i$ is given~\cite{Prosen10} in terms of the
solution $\bs{Z}$ of the Lyapunov equation $\bs{X}^T \bs{Z}+\bs{Z}\bs{X} \equiv  \bs{M}_i$ with $\bs{M}_i \equiv -\frac{i}{4} {\rm diag}\{m_{1L},\cdots,m_{KL},
\bs{0}_{1\times N},\, m_{1R},\cdots,m_{KR}\}
\otimes \bs{\sigma_y}$ and $m_{k\alpha}=\gamma\{ 2F_\alpha (\varepsilon_k)-1 \}$ as follows: Consider the change of variables $w_{2j-1}\equiv d_j+d_j^\dag,\ \ w_{2j}\equiv i(d_j-d_j^\dag)$, the NESS average of the quadratic observable $w_j w_k$ is determined by the matrix $\bs{Z}$ through the relation $\ave{ w_j w_k}=\delta_{j,k} -4i\bs{Z}_{j,k}$.
Wick's theorem can be used to obtain expectations of
higher-order observables and in fact, the full probability distribution for these
expectation values in some cases~\cite{norditaUs}.

As shown in~\cite{Prosen10} if there is no rapidity with zero real part the NESS is unique. We verify this in our model and thus all the steady state properties that we describe bellow apply to the unique NESS of the system 
%%%%%%%%%%%%%%%%%%%%%%%%%%%%%%%%%%%%%%%%%%%%%%%%%%%%%%%%%%%%%%%%%%%%%%

\subsection{Particle current} 

In this model, particle current is conserved for all parameters~\cite{mariborg}. Heisemberg equation of motions for the occupations on the mesoreservoirs $n_\alpha(\epsilon_k)=a_{k\alpha}^\dagger a_{k\alpha}$ gives~\cite{mariborg}
\begin{equation}
\frac{d}{dt}n_\alpha(\epsilon_k)=-j_\alpha(\epsilon_k)+{\mathcal D}(n_\alpha(\epsilon_k) )
\label{occalpha}
\end{equation}
with ${\mathcal D}(n_\alpha(\epsilon_k) )=-2\gamma(n_\alpha(\epsilon_k)-f_\alpha(\epsilon_k){\rm I})$ being
interpreted as the current flowing from the $\alpha$ superreservoir to the $k$ level of the $\alpha$ mesoreservoir and
$j_\alpha(\epsilon_k)=iv_{k\alpha}(a_{k\alpha}^\dagger c -c^\dagger a_{k\alpha})$ being the current flowing from the $k$ level of the $\alpha$ mesoreservoir to the quantum dot or quantum chain considering $c=c_1$ if $\alpha=L$ and $c=c_n$ if $\alpha=R$.

In the dot $\frac{d}{dt}n=J_{L}+J_{R}$ with $J_L=\sum_k j_L(\epsilon_k)$ and $J_R=\sum_k j_R(\epsilon_k).$ 
And along the chain 
\begin{equation}
\frac{d}{dt}n_l=J_{l-1}-J_l \quad l=1,\ldots,n
\end{equation}
where $J_l=-it(c^\dagger_jc_{l+1}-c^\dagger_{l+1}c_{l})$ for $l=1,\ldots,n-1$, $J_0=J_L$ and $J_n=-J_R.$

In a steady state where all time derivatives vanishes we can see that $\sum_k\langle j_L(\epsilon_k)\rangle+\sum_k\langle j_R(\epsilon_k)\rangle=0$ for the quatum dot and 
$\sum_k\langle j_L(\epsilon_k)\rangle=\langle J_0\rangle=\cdots=\langle J_n\rangle=-\sum_k\langle j_R(\epsilon_k)\rangle$ for the quantum chain. Thus the total average particle current is conserved and we can forget the chain sub-index  and call the total average particle current $\langle J \rangle$. In our model generically $\langle j_L(\epsilon_k)\rangle+\langle j_R(\epsilon_k)\rangle\neq 0$. The equality is only found in a parameter region that we call the Landauer parameter region where transport is coherent as when collisions are elastic. When dissipation ($\gamma$) is strong, the particle current is redistributed in different energy shells analogously as in inelastic scattering processes. Therefore, as we showed analytically in~\cite{mariborg}, energy current is not generically conserved but total particle current is conserved. 

%%%%%%%%%%%%%%%%%%%%%%%%%%%%%%%%%%%%%%%%%%%%%%%%%%%%%%%%%%%%%%%%%%%%%%

\subsection{Particle current in the coherent regime: The Landauer Formula}
\label{Landauer} 
The Landauer formula~\cite{DattaBook} provides an almost explicit expression for the NESS (assumed to exist and being unique) average current in a coherent system as a function of the parameters of the system. In units where $e=1$ and $\hbar=1$ it reads
\begin{equation}
\label{eq.landauer}
\ave{J}=\frac{1}{2\pi}\int d\varepsilon(f_L(\varepsilon)-f_R(\varepsilon))T(\varepsilon),
\end{equation}
where 
%\begin{equation}
$
T(\varepsilon)=
{\rm tr}[\Gamma_L(\varepsilon)G^+ (\varepsilon)\Gamma_R(\varepsilon) G^-(\varepsilon)]\,
$%\end{equation}
is the transmission probability written here in terms of 
\begin{equation}
G^\pm(\varepsilon)=\frac{1}{\varepsilon-H_S-\Sigma^\pm_L(\varepsilon)-\Sigma^\pm_R(\varepsilon)}, 
\label{green}
\end{equation}
the retarded and advanced Green function of the system connected to the leads and of $-\Gamma_\alpha/2$ the imaginary part of the self-energy $\Sigma^+_\alpha$.
For the quantum dot with symmetric coupling at the left and right mesoreservoir~\cite{Kosov11}
\begin{equation} 
T(\varepsilon)=\frac{\Gamma(\varepsilon)^2}{(U-\varepsilon+2\Lambda(\varepsilon))^2+\Gamma(\varepsilon)^2}
\label{TT}
\end{equation}
with $v_0^2\rho(\varepsilon)=\Gamma(\varepsilon)/2\pi$ and 
\begin{equation}
\Lambda(\varepsilon)=\frac{1}{2\pi}{\rm P}\int\frac{\Gamma(\omega) d\omega}{\varepsilon-\omega}=\frac{v_\alpha^2}{\theta_0} \ln\left|\frac{\varepsilon-E_{\rm min}}{\varepsilon-E_{\rm max}}\right|.
\end{equation}
In the so called wide-band limit $E_{\rm min}\ll U$ and $U\ll E_{\rm max}$,  $ \Lambda(\omega)$ can be neglected.
An explicit, but cumbersome expression of $T(\varepsilon)$ for the the quantum chain~\cite{silly,norditaUs} can be obtained. 
We will use it for the numerical examples reported bellow but we do not reproduce it. 

\subsection{Mesoreservoir occupation distribution}
Let us consider Eq.(\ref{occalpha}) for the left mesoreservoir in the NESS
\begin{equation}
\langle j_L(\varepsilon_k)\rangle=-2\gamma[\langle n_L(\varepsilon_k)\rangle-f_L(\varepsilon_k)]
\label{Key}
\end{equation}
This equation nicely shows that current, a fingerprint of the non-equilibrium steady state, can flow only if the mesoreservoir distribution departs from the equilibrium distribution. We can rewrite this equation as 
\begin{equation}
\langle n_L(\varepsilon_k)\rangle=f_L(\varepsilon_k)-\frac{1}{2\gamma}\langle j_L(\varepsilon_k)\rangle
\label{nLNESS}
\end{equation}
which gives the NESS distribution on the left mesoreservoir if the current $\langle j_L(\varepsilon_k)\rangle$ is known. When transport is coherent an analytic expression for $\langle j_L(\varepsilon_k)\rangle$ can be given. In the general case, we resort to numerical computations. 
In fact, in the coherent case $\langle j_L(\varepsilon_k)\rangle=-\langle j_R(\varepsilon_k)\rangle$ and thus $\langle j_L(\varepsilon_k)\rangle$ can be directly read from Landauer formula Eq.(\ref{eq.landauer}) for the current 
\begin{equation}
\rho_L(\varepsilon_k)j_L(\varepsilon_k)=\frac{T(\varepsilon_k)}{2\pi}[f_L(\varepsilon_k)-f_R(\varepsilon_k)].
\label{eq.landLoc}
\end{equation}
From Eq.(\ref{nLNESS}) and Eq.(\ref{eq.landLoc}) we obtain 
\begin{equation}
n_L(\varepsilon_k)=f_L(\varepsilon_k)-\frac{T(\varepsilon_k)}{4\pi\gamma \rho_L(\varepsilon_k)}[f_L(\varepsilon_k)-f_R(\varepsilon_k)]
\label{nleft}
\end{equation}
and similarly
\begin{equation}
n_R(\varepsilon_k)=f_R(\varepsilon_k)+\frac{T(\varepsilon_k)}{4\pi\gamma \rho_R(\varepsilon_k)}[f_L(\varepsilon_k)-f_R(\varepsilon_k)]
\label{nright}
\end{equation}
which is one of the central results of this paper. For the sake of generality we consider in these formulas the generic case with different density of states in the left $\rho_L(\varepsilon_k)$ and right $\rho_R(\varepsilon_k)$ mesoreservoirs but as we mentioned before in this work we consider them equal ($\rho_L=\rho_R=1/\theta_0$).
We remark that Eqs.(\ref{nleft},\ref{nright}) are valid if the current is given by Landauer formula, a situation that we discuss next with the help of numerical results.

%%%%%%%%%%%%%%%%%%%%%%%%%%%%%%%%%%%%%%%%%%%%%%%%
\begin{figure}
 \includegraphics[width=7cm]{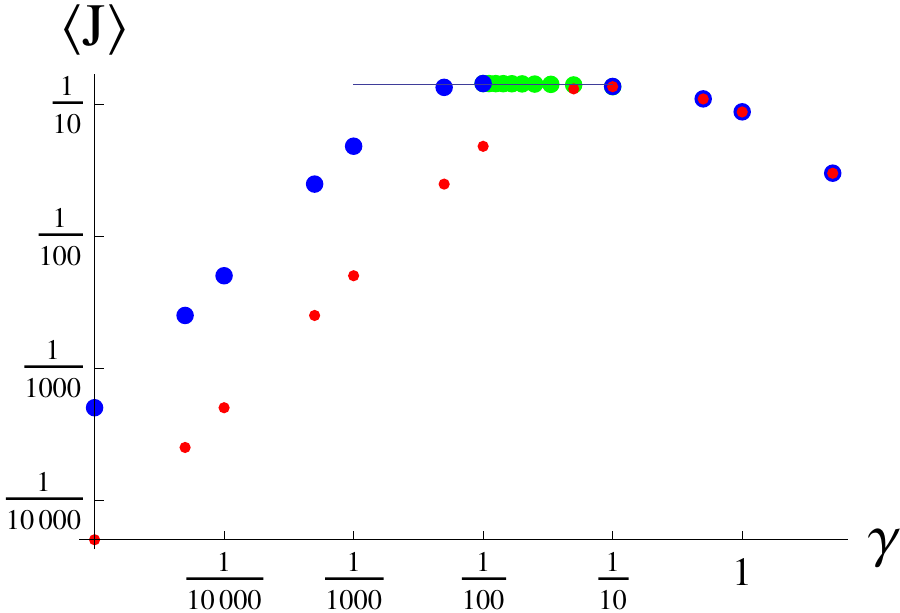}
\caption{Particle current versus $\gamma$ for the quantum dot with $U=0$, $\beta_L=5, \beta_R=10$ , 
$\mu_L=0.5,\ \mu_R=-0.5$. The red dots are for $K=51$, the blue dots are for $K=501$ and the green dots are for $\gamma=\theta_0/2=(E_{\rm max}-E_{\rm min})/(2K)$. The thin line indicates the value of the current predicted by Landauer formula for the same parameters $\beta_L, \beta_R$, $\mu_L,\ \mu_R$ and $v_0$. In this and al subsequent figures we take: $v_0=\sqrt{\theta_0/(2\pi)}$ and $E_{\rm max}=-E_{\rm min}=5$.}
\label{j-gamma}
\end{figure}
% and $E_{\rm max}=-E_{\rm min}=20$
%%%%%%%%%%%%%%%%%%%%%%%%%%%%%%%%%%%%%%%%%%%%%%%

%%%%%%%%%%%%%%%%%%%%%%%%%%%%%%%%%%%%%%%%%%%%%%%%%%%%%%%%%%%%%
\begin{figure}
\includegraphics[width=7cm]{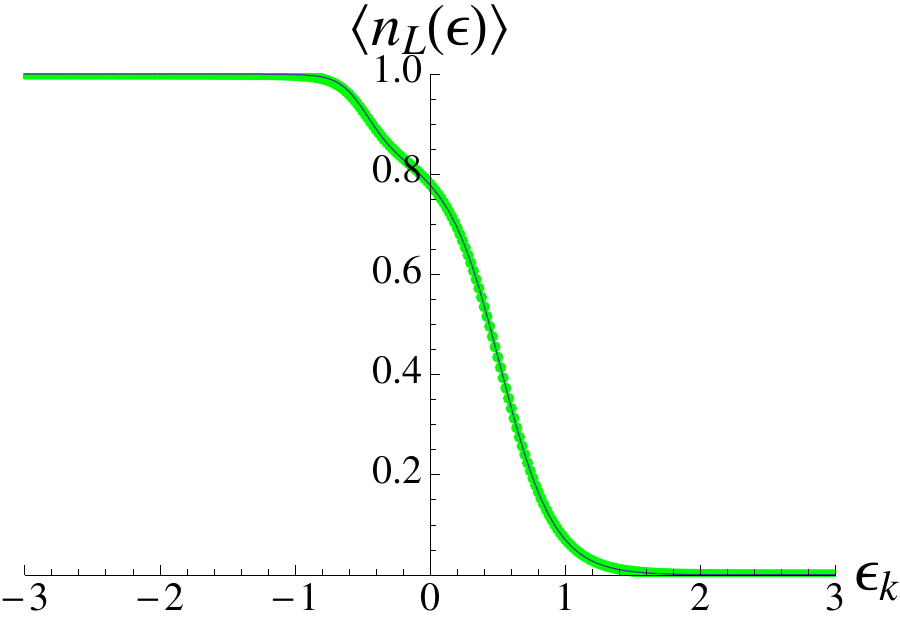}
\includegraphics[width=7cm]{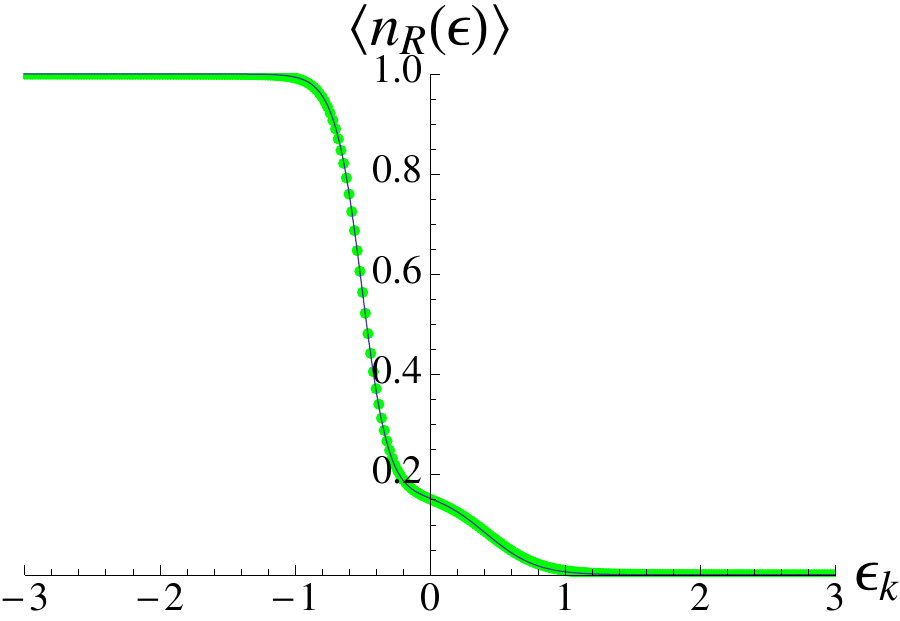}
\caption{Mesoreservoirs occupation distributions in the Landauer regime. The dots are computed numerically for the single dot model in the NESS
a) Green dots: occupation distributions for the left mesoreservoir with $K=500$. Continuous line is obtained from Eq.(\ref{nleft})
with parameters chosen according to Eq.(\ref{landparameterreg}). 
b) Green dots: occupation distributions for the right mesoreservoir with $K=250$. Continuous line is obtained from Eq.(\ref{nright})
with parameters chosen according to Eq.(\ref{landparameterreg}). }
\label{occup}
\end{figure}
%%%%%%%%%%%%%%%%%%%%%%%%%%%%%%%%%%%%%%%%%%%%%%%%%%%%%%%%%%%%%%%
\subsection{Numerical results}
\label{numeric}

{\it The current}:
In Fig.\ref{j-gamma} we plot the total particle current $\langle J\rangle$ through the dot as a function of $\gamma$ for different values of $K$. For a fix value of $K$, we observe an initial linear growth of $\langle J\rangle$ with $\gamma$ 
consistent with the fact that for small $\gamma$ the super-reservoirs are weakly coupled to the mesoreservoirs plus dot system and therefore the distribution $n_L$ differs from $f_L$ and according to Eq.(\ref{Key}) the current grows linearly with $\gamma$. We also observe that if we consider a fix small value of $\gamma$ in this linear regime, $\langle J\rangle$ increases if we increase $K$ (compare big blue dots with small red dots) but never goes beyond the value computed from Landauer formula Eq.(\ref{eq.landauer}) with $T(\varepsilon)$ from Eq.(\ref{TT}),indicated by the thin line in Fig.\ref{j-gamma}. In fact we observe that when the current reaches the Landauer value it stays there in a {\it plateau} and then starts to decay with $\gamma$. Interestingly the curves 
obtained for different $K$ superpose in this decaying regime. This provides interesting information about the strong coupling regime. For large $\gamma$ the mesoreservoir distribution approaches the corresponding Fermi distribution and from Eq.(\ref{Key}) we can conclude that $n_L-f_L\sim{\mathcal O}(\gamma^{-1-\alpha}/K)$. The  {\it plateau} occurs at the maximum value of the current which is moreover given by the  value predicted by Landauer formula. We observe that 
by choosing $\gamma$ and $K$ such that    
%L A N D A U E R  P A R A M E T E R  R E G I O N
%
\begin{equation}
2\gamma\rho_\alpha=2\gamma/\theta_0\approx 1,
\label{landparameterreg}
\end{equation} 
the computed value of the current always fall on the Landauer value (green dots). Condition Eq.(\ref{landparameterreg}) 
defines the Landauer parameter region mentioned above and is interpreted as 
equating the width (due to contact with environment) of each level of the mesoresevoir $2\gamma$  with the level spacing $\theta_0$. In other words, the density of states in the mesoreservoir is smooth.  We have also check numerically (data not shown) that in this line ($\gamma=\theta_0/2$) the current is conserved on shell i.e. $j_L(\varepsilon_k)=-j_R(\varepsilon_k)$, which is consistent  with the fact that particle current is given by Landauer formula. 

\begin{figure}
\includegraphics[width=7cm]{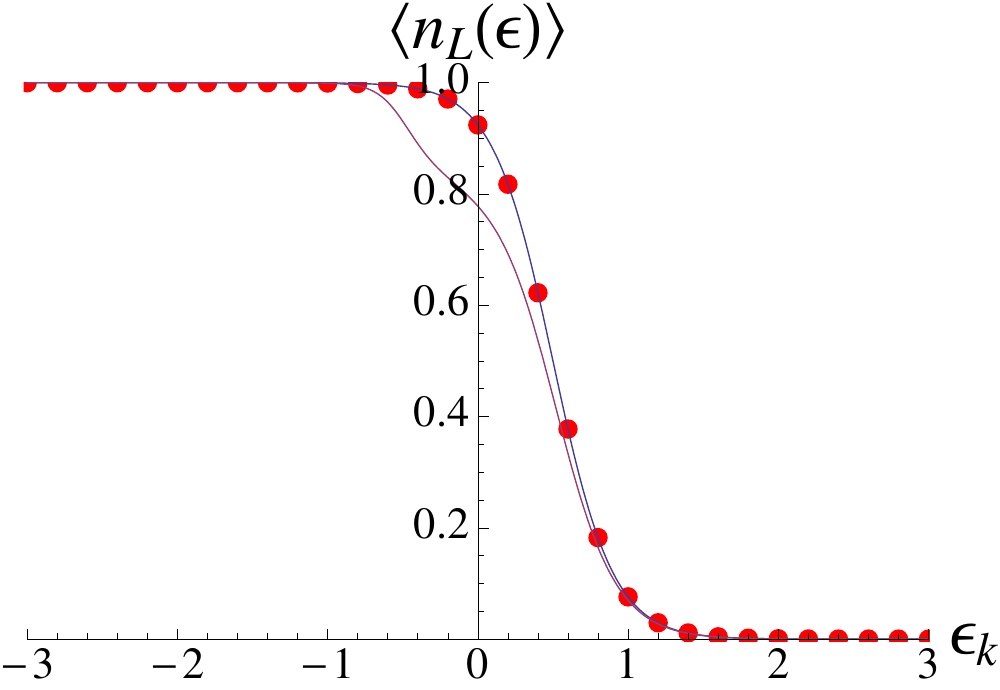}
\includegraphics[width=7cm]{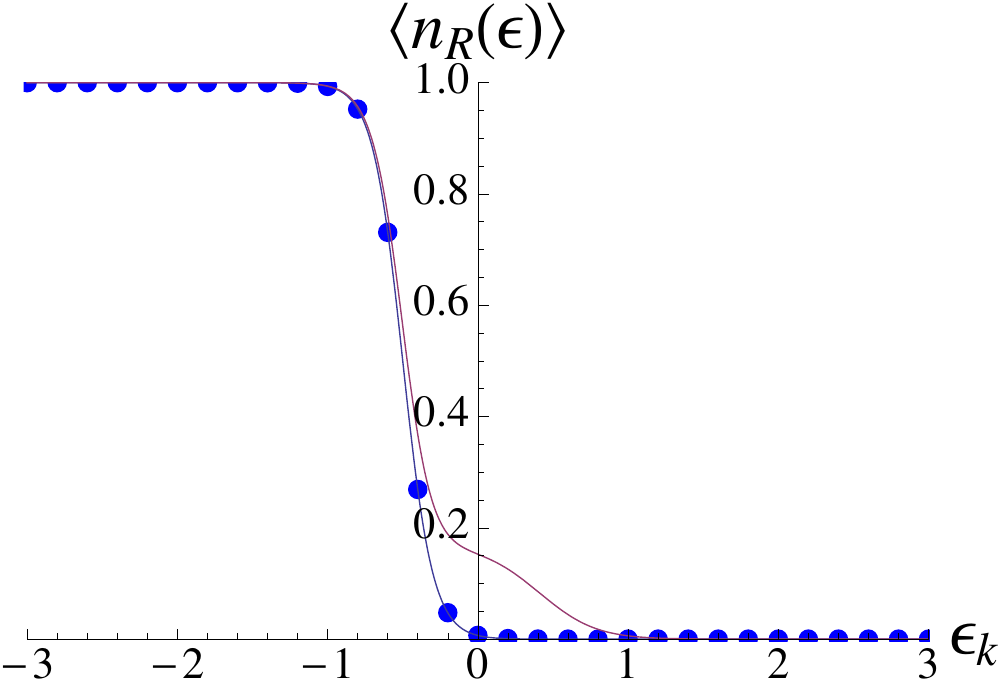}
 
\caption{Mesoreservoirs occupation distributions in the strong coupling limit $\gamma=10$.
The dots are computed numerically for the single dot model in the NESS with  $K=51$. 
a) Red dots: occupation distributions for the left mesoreservoir. Continuous line under the dots is $f_L(\epsilon)$ while the red line is 
the expected population in the Landauer regime obtained from Eq.(\ref{nleft})
with parameters chosen according to Eq.(\ref{landparameterreg}). 
b) Blue dots: occupation distributions for the right mesoreservoir. Continuous line under the dots is $f_R(\epsilon)$ while the red line is 
the expected population in the Landauer regime obtained from Eq.(\ref{nright})
with parameters chosen according to Eq.(\ref{landparameterreg}). 
}
\label{occupstrong}
\end{figure}

\begin{figure}
\includegraphics[width=7cm]{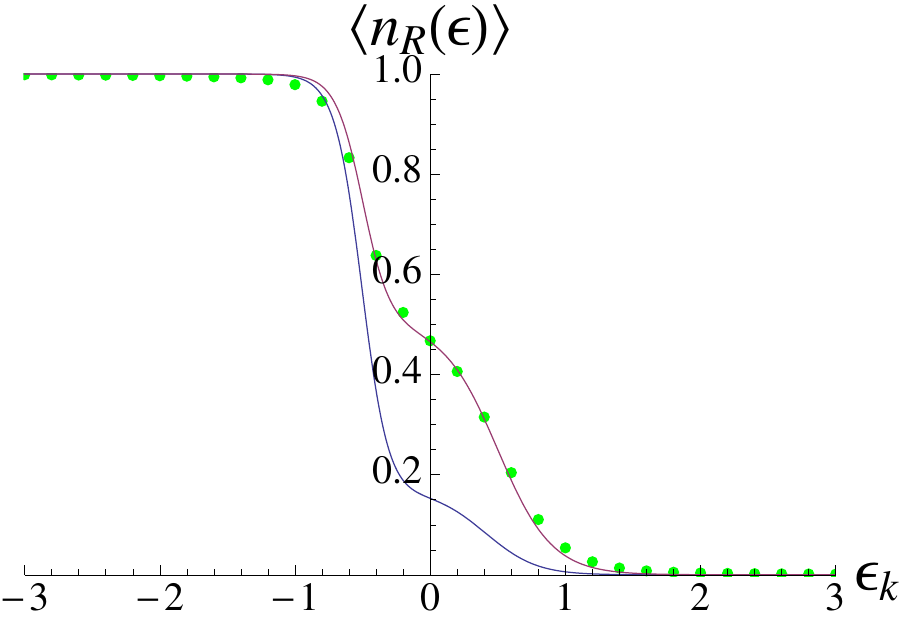}
 
\caption{Green dots: mesoreservoirs occupation distribution in the weak coupling regime for the right mesoreservoir for $\gamma=10^{-5}$ and $K=51$.
The continuous line under the dots is given by $(f_L+f_R)/2$, while the blue line is the expected population in the Landauer regime. The left distribution is identical to the right distribution and we do not show it.}
\label{occupweak}
\end{figure}

{\it Occupation}:
Now we analyze the mesoreservoir occupation distribution for the different regimes of $\langle J\rangle$ versus $\gamma$ we just discussed. 
\begin{itemize}
\item In the strong dissipation regime $\gamma>\theta_0/2$, where the current start to decay as $\gamma$ increases (see Fig.\ref{j-gamma}), we expect that $n_\alpha(\epsilon_k)\approx f_\alpha(\epsilon_k)$
as is verified in Fig. \ref{occupstrong}

\item In the weak dissipation regime $\gamma<\theta_0/2$, where the current grows linearly with $\gamma$ (see Fig.\ref{j-gamma}), the super reservoir are almost decoupled from the mesoreservoir and the whole mesoreservoir-dot system is isolated. In that case, we expect from non-equilibrium Green function formalism that the occupation of the diagonal modes of the full Hamiltonian $H=H_L+H_S+H_R+V$ is the average of left and right Fermi distribution. Thus for large mesoreservoirs (large $K$) we expect
$n_L(\epsilon_k)=n_R(\epsilon_k)\approx (f_L(\epsilon_k)+f_R(\epsilon_k))/2$ in the symmetric case considered, as verified in Fig.\ref{occupweak}.

\item In the Landauer regime i.e. when $2\gamma/\theta_0\approx 1$ and the current is in the  {\it plateau}, the non-equilibrium distribution is predicted by Eq.(\ref{nleft}) with $2\gamma\rho_L=1$ i.e.,
\begin{equation}
n_L(\varepsilon_k)=f_L(\varepsilon_k)-\frac{T(\varepsilon_k)}{2\pi}[f_L(\varepsilon_k)-f_R(\varepsilon_k)]
\label{nllandauer}
\end{equation}
 which is plotted in Fig.\ref{occup} and perfectly agree with the numerical data.

\end{itemize}
Note  that the first two regimes are not physically relevant for our context here, which is to model the current and the leads in a typical non-equilibrium
situation where the parameters $\gamma$ and $K$ are not under control.

%%%%%%%%%%%%%%%%%%%%%%%%%%%%%%%%%%%%%%%%%%%%%%%%%%%%%%%%%%%%%%%%%
\begin{figure}
\includegraphics[width=7cm]{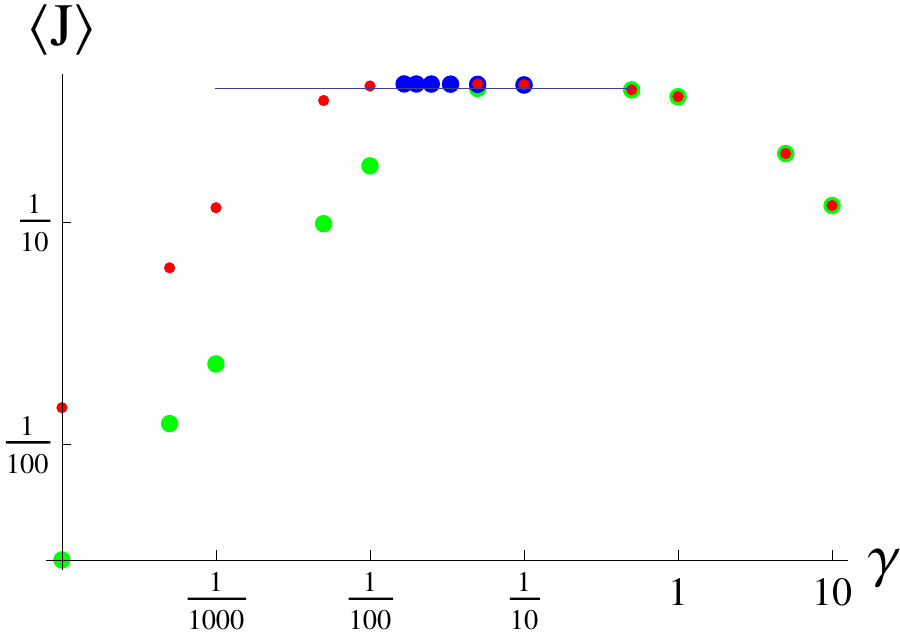}
\includegraphics[width=7cm]{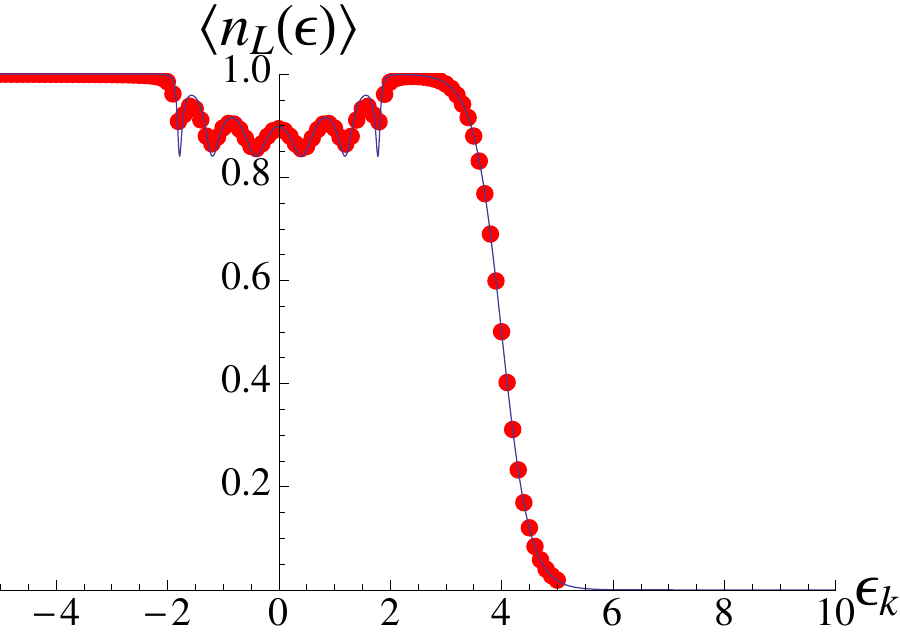}
\caption{For the chain with $N=6$, $K=101$, $t_j=1,U_j=0$, $\beta_L=\beta_R=4$, $\mu_L=-\mu_R=4$  a) current  and b) mesoreservoirs occupation distribution in the Landauer regime. The red dots are numerical computations and the continuous line is from Eq.(\ref{nllandauer}) with $T(\varepsilon)$ computed in~\cite{norditaUs}.}
\label{chain}
\end{figure}

%%%%%%%%%%%%%%%%%%%%%%%%%%%%%%%%%%%%%%%%%%%%%%%%%%%%%%%%%%%%%%%%%%%
We have done the same analysis for a chain. The same picture is observed for particle current as a function of $\gamma$ and for the occupation of the mesoreservoirs, but since the transmission coefficient has an interesting structure as a function of energy, these structure is observed in the mesoreservoir distribution
as illustrated in Fig.\ref{chain}

Physically the distributions $n_L(\varepsilon_k)$ and $n_R(\varepsilon_k)$ on the mesoreservoirs correspond to the distribution on the leads, since mesoreservoirs model the leads that connect the superreservoirs to the system.
The non-equilibrium distribution of the leads has been considered in~\cite{Pinhas1985} and in more detail in~\cite{DattaBook}. 
We compare both results bellow.

%%%%%%%%%%%%%%%%%%%%%%%%%%%%%%%%%%%%%%%%%%%%%%%%%%%%%%%%%%%%%%%%%%%%%%

\subsection{Mesoreservoirs versus Fully coherent leads}
\label{datta}

In the Landauer picture~\cite{Pinhas1985,DattaBook} the contacts are reflectionless thus the occupation distribution of electrons traveling to the right at the left lead is $n_L^+=f_L$ while at the right lead is $n_R^+=T f_L+ R f_R=f_R+T(f_L-f_R)$. Analogously the distribution of left going electrons at the right lead is $n_R^-=f_R$ while at the left lead is $n_L^-=T f_R+R f_L=T(f_R-f_L)+f_L$. 
From the difference of the distribution at the left and right we can evaluate the
the potential drop across the scatterer (from left to right) for left and right moving electrons and obtain $\Delta \mu_c=(1-T)(\mu_L-\mu_R)$ at zero temperature.
Accordingly we can define a conductance of the barrier  $G_0$ which is related to the conductance $G$ by $G_0=\frac{I}{\Delta \mu_c}=\frac{1}{1-T}\frac{I}{\Delta \mu}=\frac{G}{R}$ as originally suggested by Landauer. 
We can also compute the potential drop from the reservoir to the lead and the associated contact resistance $1/G_c$ obtaining that $1/G_c +1/G_0+1/G_c=1/G$. 

If we consider that in the leads for every right going state there is a left going states with the same energy  then the occupation for the energy states on the left lead is $f_L+(T/2)(f_R-f_L)$ which differ from $n_L$ in Eq.(\ref{nllandauer}). 
The difference is due to the following: 
In our model, the mesoreservoirs are subject to dissipation, so in the Landauer's view they are still part of the reservoir and therefore the potential drop associated to the difference between $f_L$ and $n_L$ is less than the potential drop associated to the contact resistance. In this sense mesoreservoirs are intermediate between reservoirs (described by Fermi distribution) and leads (that are not affected by dissipation). Note that in the Landauer picture the system plus leads is considered to be coherent, thus contact resistance and system resistance are added incoherently.

%%%%%%%%%%%%%%%%%%%%%%%%%%%%%%%%
\section{conclusions}
\label{conclu}

We have studied a model of coherent transport where the non-equilibrium distribution of mesoreservoirs can be analyzed. 
In this model Landauer prediction for the current holds under certain (and well understood) choices of the parameters and it is the maximum particle current of our model. We would like to point out that it may be possible to control dissipation in the experimental setting discussed in~\cite{science}, therefore
other parameter regions (out of Landauer region) of our model could be relevant. 

The non-equilibrium distribution 
on the mesoreservoirs [Eqs. (\ref{nleft},\ref{nright}) in the parameter region defined by Eq.(\ref{landparameterreg})] differs by numerical factors form the ones discussed in section \ref{datta} expected from Landauer's picture. As we commented above, in our model dissipation act on the mesoreservoirs and therefore they are not part of the coherent system. Indeed there is a potential drop between our distribution Eq. (\ref{nleft}) 
and the one expected from Landauer's picture (see section \ref{datta}). In this sense with mesoreservoirs there is a smother potential drop between left reservoir
and conductor.

\acknowledgments

FB thanks Fondecyt 1110144, Conicyt-ANR 38 and Anillo ACT 127 projects. SA  thanks Fondecyt project 3120254. We 
also thanks discussions with C. Mejia-Monasterio, T. Prosen, M. Esposito and B. Zunkovic.

\end{document}